\documentclass[prd,preprintnumbers,showkeys,superscriptaddress,showpacs,
nofootinbib,fleqn]{revtex4-1}   

\usepackage{amsmath,amsfonts,amssymb,amscd,amsxtra,amsthm}
\usepackage{graphicx}
\usepackage{bm}
\usepackage{epstopdf}
\usepackage{color}

\begin{document}  
\preprint{INHA-NTG-09/2011}
\title{Transverse strange quark spin structure of the nucleon} 
\author{Tim Ledwig}
\email[E-mail: ]{ledwig@kph.uni-mainz.de}
\affiliation{Institut f\"ur Kernphysik, Universit\"at Mainz, D-55099
  Mainz, Germany} 
\author{Hyun-Chul Kim}
\email[E-mail: ]{hchkim@inha.ac.kr}
\affiliation{Department of Physics, Inha University, Incheon 402-751,
Korea}  
\affiliation{School of Physics, Korea Institute for Advanced Study,
  Seoul 130-722,  Korea} 
\date{October, 2011}
\begin{abstract}
We investigate the transverse quark spin densities of the nucleon with
the lowest moment within the framework of the SU(3) chiral
quark-soliton model, emphasizing the strange quark spin density. 
Based on previous results of the vector and tensor
form factors, we are able to determine the impact-parameter dependent
probability densities of transversely polarized quarks in an
unpolarized nucleon as well as those of unpolarized quarks in a
transversely polarized nucleon. We find that the present numerical
results for the transverse spin densities of the up and down quarks
are in good agreement with those of the lattice calculation. We
predict the transvere spin densities of the strange quark. It turns
out that the polarized strange quark is noticeably distorted in an
unpolarized proton.
\end{abstract} 
\maketitle

\section{Introduction}
The transversity of the nucleon is one of the most
important issues in understanding the structure of the
nucleon~\cite{Ralston:1980pp,Jaffe:1991ra,Cortes:1991ja} (see also the   
review~\cite{review,Boffi:review,Hagler:2009ni,Miller:2010nz}).  
Though it is rather difficult to get access to the transversity of the 
nucleon experimentally because of its chiral-odd nature, 
some information is now available from the transverse spin asymmetry
$A_{TT}$ of Drell-Yan processes in $p\overline{p}$
reactions~\cite{Efremov:2004,Anselimo:2004,PAX:2005,Pasquini:2006}. The
transversity distribution $\delta q(x)$ was also extracted, based on a 
global analysis of the data of the azimuthal single spin asymmetry in
SIDIS processes $lp^{\uparrow}\to l\pi X$ by the
Belle~\cite{Belle:2006}, HERMES~\cite{HERMES:2005,HERMES:2006} and
COMPASS~\cite{COMPASS:2007}  collaborations. The corresponding tensor
charges for the up and down quarks were reported~\cite{Anselmino:2009}. 
Moreover, the QCDSF/UKQCD Collaborations has announced the first
lattice results of the transverse spin structure of the
nucleon~\cite{Gockeler:2006zu} based on the work~\cite{Diehl:2005jf}. 
The work \cite{Diehl:2005jf} extended the analysis of
Refs.~\cite{Burkardt:2000za,Burkardt:2002hr} for the generalized
parton distributions (GPDs) of the vector and axial-vector current to
the generalized transversity distributions of the tensor
current. Furthermore, the Refs.~\cite{Burkardt:2000za,Burkardt:2002hr}
provide a mean to obtain information on the parton distributions in
the impact parameter space that is not affected by certain
relativistic ambiguities.   

In the present work, we investigate the transverse quark spin
densities of the nucleon with the lowest moment, based on 
the previous results for the nucleon vector and tensor form factors
obtained within the SU(3) chiral quark-soliton model ($\chi$QSM). In
particular, we emphasize the transverse spin densities of the strange
quark in the nucleon. The nucleon vector form factors were already
computed within that framework in  
Refs.~\cite{SilvaThesis,Silva:2001st,Silva:2002ej}. As a result, the
strange vector form factors were predicted and turned out to be within
the uncertainty of the experimental data~\cite{Goeke:2006gi}.   
The tensor form factors of the nucleon have been recently calculated
within the $\chi$QSM~\cite{Ledwig:2010tu}. The results turn out to be
similar to those of the lattice calculation. The strange tensor charge
of the nucleon was also predicted in Ref.~\cite{Ledwig:2010tu} and was
found to be rather small. The anomalous tensor
magnetic moments (ATMM) of the nucleon were computed in
Ref.~\cite{Ledwig:2010zq} within the same framework and were found to
be positive and large for both up and down quarks as in the lattice
calculation~\cite{Gockeler:2006zu}. The ATMM plays an 
important role in describing the transverse deformation of the
transverse polarized quark distribution in an unpolarized
nucleon. 

We use our previous results for these form factors and investigate
them when viewed from the light front moving towards
the nucleon. In this frame it is possible to extract spatial
information on the nucleon, in which certain ambiguities arising from
relativistic corrections are
absent~\cite{Burkardt:2000za,Burkardt:2002hr}. We will follow the
strategy of the
works~\cite{Diehl:2005jf,Gockeler:2006zu,ChargeDensities,    
 TiatorVanderhaeghen} in SU(2), where form factors obtained from
lattice calculations and experimental data were used to
map out nucleon transverse spin structures and transverse charge
densities. We would also like to note the comprehensive investigations
on generalized transverse momentum-dependent 
parton distributions~\cite{Lorce:2011a,Lorce:2011b} within the
$\chi$QSM and the light-front constituent quark model, both in the 3 quark
valence approximation. However, since we employ the SU(3) version of
the $\chi$QSM with explicit breaking of SU(3) symmetry  together with
the whole Dirac sea, we are in a unique position to obtain spatial
information on the transverse spin sturcture of the strange quark
inside the 
nucleon.  

\section{Transverse quark spin densities}
\textbf{2.} The vector and tensor form factors of the nucleon
are defined in terms of the matrix elements of the vector and tensor
currents, respectively: 
\begin{eqnarray}
\langle
N_{s^{\prime}}(p^{\prime})| \overline{\psi}(0) \gamma^{\mu}
\lambda^\chi \psi(0)|N_{s}(p) \rangle & = &
\overline{u}_{s^{\prime}}(p')
\left[F_{1}^\chi\left(Q^{2}\right) \gamma^{\mu} + F_{2}^\chi
  \left(Q^{2} \right) \frac{i 
    \sigma^{\mu\nu}q_{\nu}}{2M_N} \right] u_{s}(p)\,,
\cr
\langle N_{s^{\prime}}(p^{\prime})| \overline{\psi}(0) i
\sigma^{\mu\nu} \lambda^\chi \psi(0)|N_{s} (p)\rangle & = &
\overline{u}_{s^{\prime}}(p^{\prime}) \left[H_{T}^\chi (Q^{2})
  i\sigma^{\mu\nu}+E_{T}^\chi (Q^{2}) \frac{\gamma^{\mu} 
    q^{\nu}-q^{\mu} \gamma^{\nu}}{2M_N}  \right. \cr
&& \left. \hspace{1.96cm} +\,\tilde{H}_{T}^\chi (Q^{2})
  \frac{(n^{\mu}q^{\nu}  - q^{\mu}n^{\nu})}{2M_N^{2}}\right]u_{s}(p)\,,
\label{eq:vtform}
\end{eqnarray}
where $\gamma^\mu$ denotes the Dirac matrix and $\sigma^{\mu\nu}$ is the
spin operator defined as $\sigma^{\mu\nu} =
i[\gamma^\mu,\,\gamma^\nu]/2$. The $\lambda^\chi$ represent the
Gell-Mann matrices including the unit matrix $\lambda^0 =
\sqrt{2/3}\mathbf{1}$. The $\psi$ stands for the quark field and
$u_s(p)$ designates the spinor for the nucleon with mass $M_{N}$,
momentum $p$, and the third component of its spin $s$. The momentum
transfer $q$ and the total momentum $n$ are defined as $q=p'-p$ with
$Q^2=-t=-q^2$ and $n=p+p'$. The form factors in Eq.(\ref{eq:vtform})
are related to the generalized form factors defined in 
Ref.~\cite{Hagler:2003,Diehl:2005jf} as follows: $F_1^\chi(Q^2) =
A_{10}^\chi(t)$, $F_2^\chi(Q^2) = B_{10}^\chi (t)$,
$H_{T}^{\chi}(Q^2)=A_{T10}^\chi(t)$,
$E_T^\chi(Q^2)=B_{T10}^\chi(t)$, and
$\tilde{H}_T^\chi(Q^2)=\tilde{A}_{T10}^\chi(t)$. Note that also the 
GPDs $H(x,0,t)$, $E(x,0,t)$, $H_T(x,0,t)$, $E_T(x,0,t)$ and
$\tilde{H}_T(x,0,t)$ are related to the form factors $F_1(t)$,
$F_2(t)$, $H_T(t)$, $E_T(t)$ and $\tilde{H}_T(t)$ with the momentum
fraction $x$ integrated. We use the arguments of a given function to
distinguish between GPDs, form factors and Fourier transformed form
factors. Furthermore, we define the anomalous magnetic moment $\kappa$
and the tensor anomalous magnetic form factor as: 
\begin{equation}
\kappa \;=\; F_2(0),\;\;\;\;
\kappa_{T}\left(Q^2\right)=E_{T}(Q^2)+2\tilde{H}_{T}(Q^2),
  \label{eq:kappa}
\end{equation}
with $\kappa_T = \kappa_T(0)$ as the ATMM. We want to mention that
$\kappa_T$ is a more important quantity than the two form factors
$E_{T}$ and $\tilde{H}_{T}$, since it is involved directly in
describing the spatial distribution of the nucleon in the transverse
plane.  

The transverse spin densities of quarks with transverse polarization $\bm
s$ in a nucleon with transverse spin $\bm S$ are expressed as
\cite{Diehl:2005jf} 
\begin{eqnarray}
\rho(\bm b,\,\bm S,\, \bm s) & = & \frac{1}{2}\left[\,\,
H(b^2)-S^{i}\epsilon^{ij}b^{j}\frac{1}{M_{N}}
\frac{\partial E (b^2)}{\partial b^2} -s^{i}\epsilon^{ij}b^{j}\frac{1}{M_{N}}
\frac{\partial\kappa_{T} (b^2)}{\partial b^2} \right.\cr
&  &
\left. +s^{i}S^{i}\left\{H_{T}(b^2) - \frac{1}{4M_{N}^{2}}
\nabla^2\tilde{H}_{T}(b^2)
\right\}+s^{i} \left( 2b^{i}b^{j}-b^{2}  \delta^{ij}\right) 
S^{j}\frac{1}{M_{N}^{2}}\left(\frac{\partial}{\partial b^2}\right)^2
\tilde{H}_{T} (b^2) \right]\,,
\label{eq:rho}
\end{eqnarray}  
where $\bm b$ denotes the two dimensional vector in impact parameter
space with distance $b=\sqrt{b_{x}^{2}+b_{y}^{2}}$ from the center
of the nucleon momentum. The tensor $\epsilon^{ij}$ is an 
antisymmetric tensor with the property
$\epsilon^{12}=-\epsilon^{21}=1$.  The operator $\nabla^2$ is
a Laplacian with respect to $\bm b$.
The $b^2$-dependent form factors are given by
the Fourier transformations of the vector and tensor form
factors which are written generically as 
\begin{equation}
 \mathcal{F}^\chi (b^2) = \int^\infty_0 \frac{dQ}{2\pi} Q J_0(bQ)F^\chi(Q^2), 
\end{equation}
where $J_0$ is a Bessel-function with order zero. 
Note that the transverse quark spin densities in Eq.(\ref{eq:rho}) are
just the first moments of the two dimensional spin densities 
$\rho(x,\,\bm b, \bm S,\bm s)$ that indicate the probability 
of finding a quark with momentum fraction $x$ and transverse
polarization $\bm s$ at distance $b$ in a nucleon with polarization
${\bm S}$~\cite{Diehl:2005jf,Gockeler:2006zu}.  

\section{Flavor vector and tensor form factors of the nucleon} 
The vector and tensor form factors defined in
Eq.(\ref{eq:vtform}) have been already studied in
detail~\cite{SilvaThesis,Silva:2001st, Silva:2002ej,
Ledwig:2010tu,Ledwig:2010zq}.  Since they are the basis of the present work, 
we will briefly recapitulate the 
results. We refer to Ref.~\cite{Christov:1995vm} for a general
formalism as to how to compute form factors within the  
$\chi$QSM. There are four different parameters in the $\chi$QSM: the
cut-off mass $\Lambda$, the dynamical quark mass $M$, the current
quark mass of the up and down quark $\bar{m}$, and the strange current
quark mass. Apart from the dynamical quark mass $M$, all other
parameters are fixed in the mesonic sector by using the pion decay
constant $f_\pi=93$ MeV, physical pion mass $m_\pi=139$ MeV, and the
kaon mass. Since various observables of the nucleon such as the
electric charge radii are well reproduced with the value of $M=420$
MeV, we will use all the results of the relevant form 
factors produced with this value.

The vector and tensor form factors $\mathcal{F}^q$ for the up,
down and strange quarks ($q=u$, $d$, $s$) can be expressed in terms of
flavor form factors $\mathcal{F}^\chi$ with $\chi=0,\,3,\,8$:
\begin{eqnarray}
\mathcal{F}^{u} & = &
\frac{1}{2}\left(\frac{2}{3}\mathcal{F}^{0}+\mathcal{F}^{3} +
  \frac{1}{\sqrt{3}} \mathcal{F}^{8}\right),\cr
\mathcal{F}^{d} & = &
\frac{1}{2}\left(\frac{2}{3} \mathcal{F}^{0} - \mathcal{F}^{3} +
  \frac{1}{\sqrt{3}} \mathcal{F}^{8}\right),\cr
\mathcal{F}^{s} & = & \frac{1}{3}\left(\mathcal{F}^{0} - \sqrt{3}
  \mathcal{F}^{8} \right)\,.
\label{eq:flavor} 
\end{eqnarray}

The vector form factors have been already studied in
Refs.~\cite{SilvaThesis,Silva:2001st, Silva:2002ej}. In particular,
the strange vector form factors were predicted and turned out to be
within the uncertainty range of the experimental
data~\cite{Goeke:2006gi}. 

\begin{figure}[ht]
\includegraphics[scale=0.43]{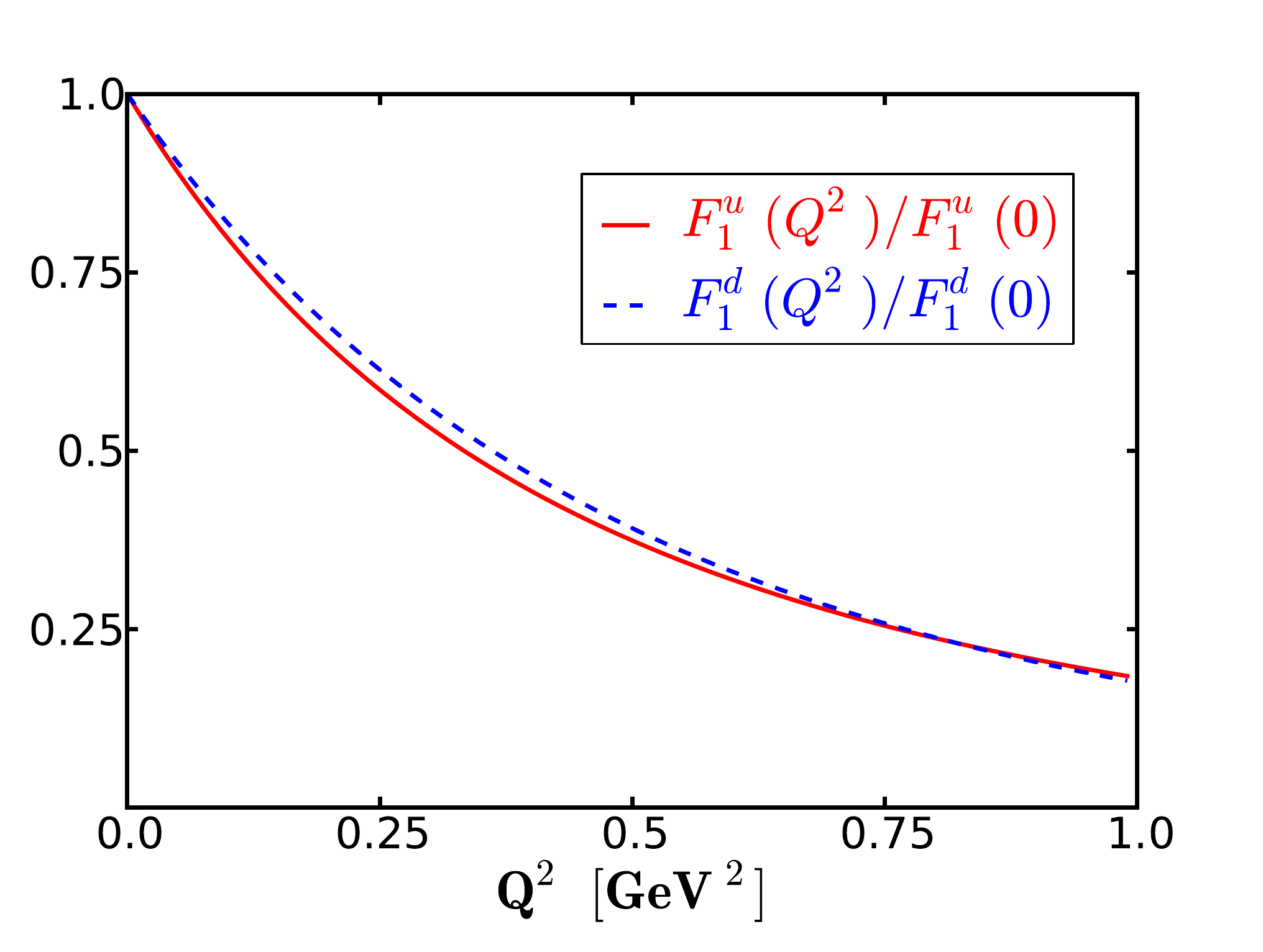}\includegraphics[scale=0.43]{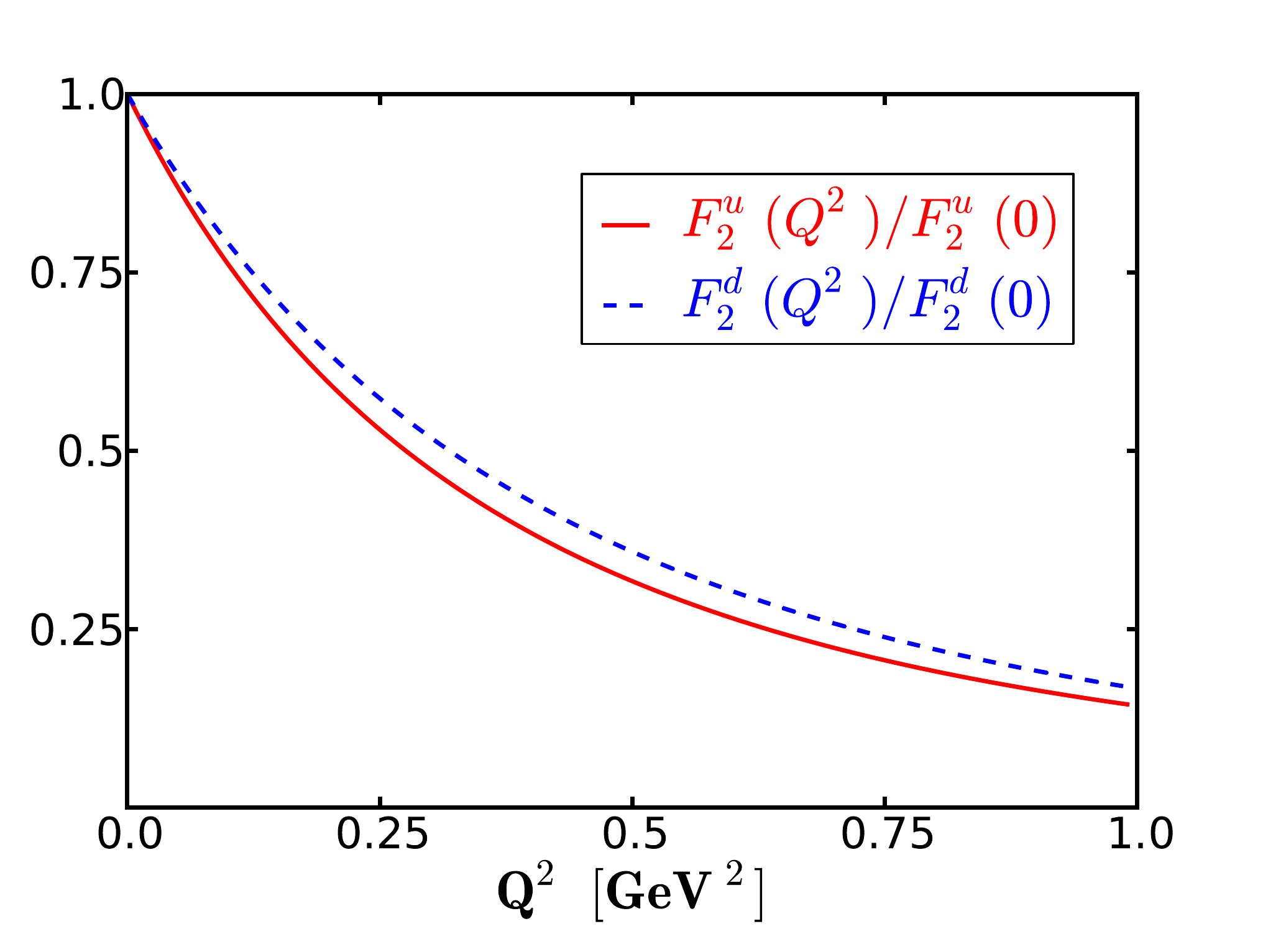} 
\caption{Vector form factors of the proton for the up and down quarks
  from the $\chi$QSM. The values at $Q^{2}=0$ for excluding the charge but
  including the number of valence quarks are given as
  $F_{1}^{u}\left(0\right)=2$, $\kappa^u=1.35$,
  $F_{1}^{d}\left(0\right)=1$, and $\kappa^d=-1.80$. The 
  anomalous magnetic moment of the proton is obtained as
  $\kappa_{p}=1.49$ while the experimental value is  
$\kappa_{p}^{\mathrm{Exp}}=1.79$.} 
\label{fig:1}
\end{figure}

\begin{figure}[ht]
\includegraphics[scale=0.6]{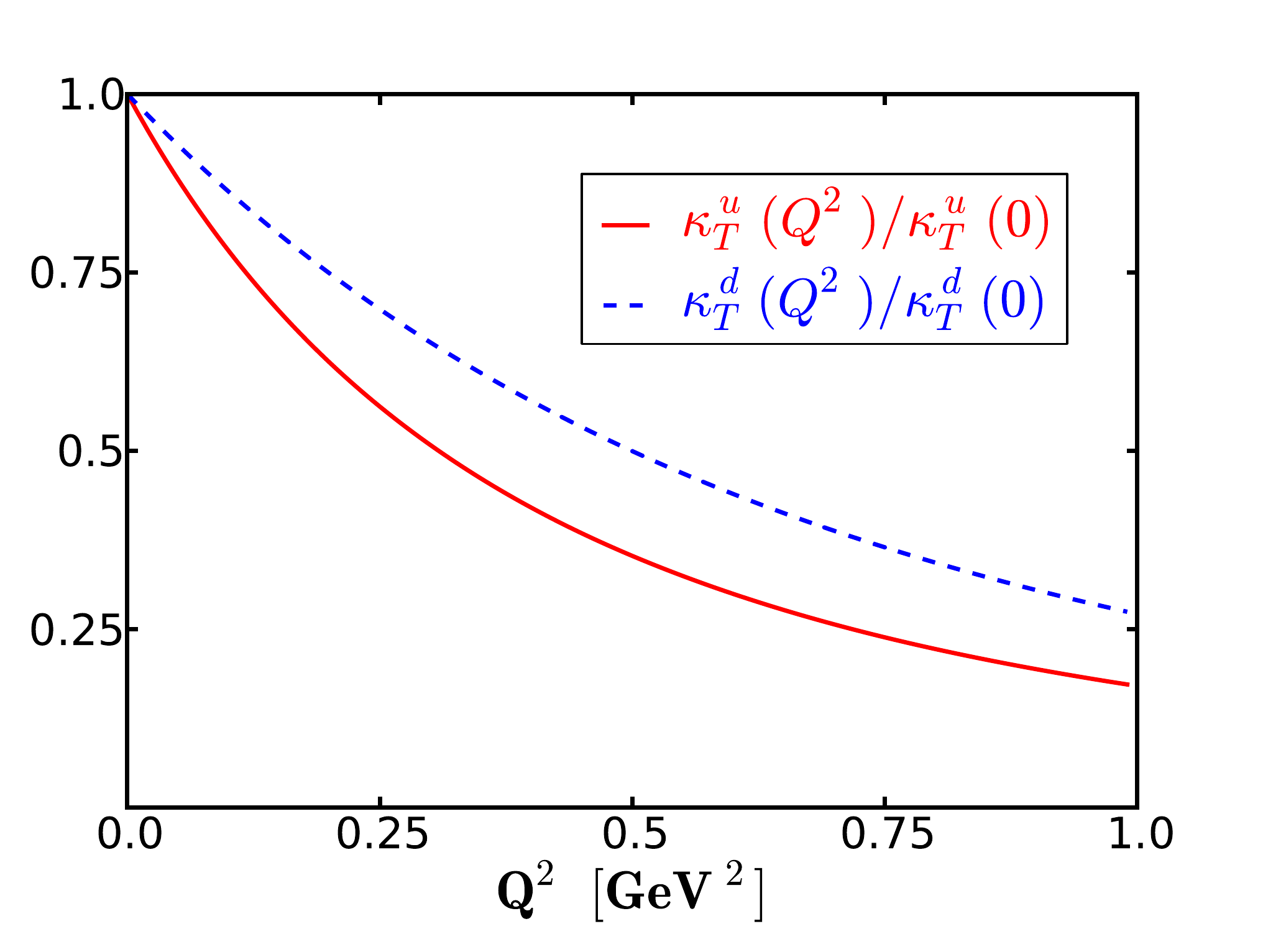}
\caption{Anomalous tensor magnetic form factors of the proton for the 
up and down quarks from the $\chi$QSM. The values at
$Q^{2}=0$ are given as $\kappa_{_{T}}^{u}=3.56$ and
$\kappa_{T}^{d}=1.83$. The values of the lattice calculation are
$\kappa_{_{T}}^{u}=3.70$ and $\kappa_{T}^{d}=2.35$ at the scale of 
$\mu^2=0.36\,\mathrm{GeV}^2$~\cite{Gockeler:2006zu}.} 
\label{fig:2}
\end{figure}

\begin{figure}[ht]
\includegraphics[scale=0.6]{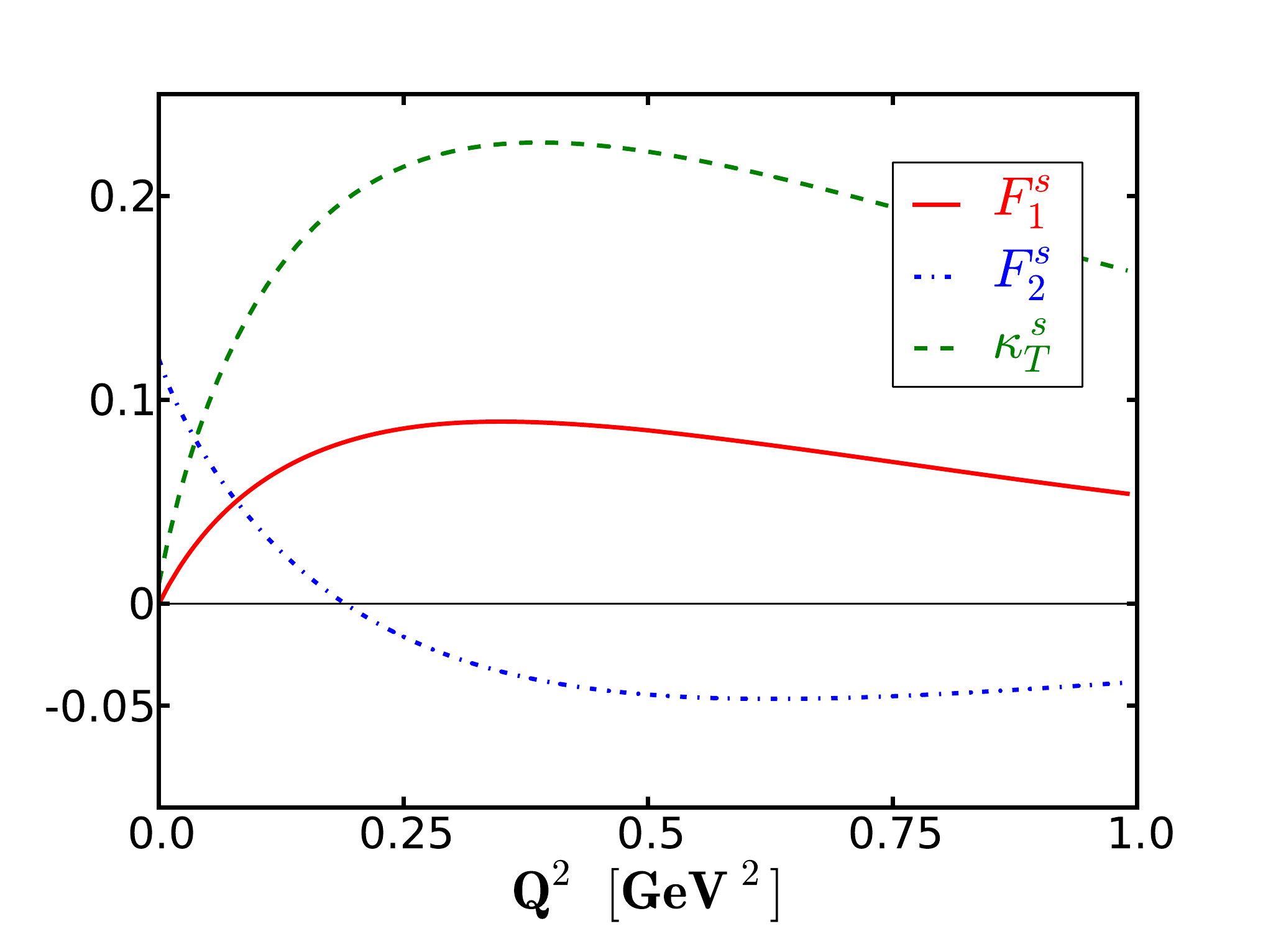}
\caption{Strange vector and anomalous tensor magnetic form factors of
  the proton from the $\chi$QSM. The charges are excluded for the
  vector form factors where the strange magnetic moment of the
  proton is obtained as $\mu_{s}=0.12\,\mu_{N}$.}  
\label{fig:3}
\end{figure}

Figure~\ref{fig:1} draws the proton up and down Dirac form
factors as a function of $Q^2$ in solid and dashed curves,
respectively. Note that they are normalized to one at $Q^2=0$ in order
to compare their $Q^2$ dependences and look very similar to each other.
Fig.~\ref{fig:2}, the up and down Pauli form factors are depicted,
which are also normalized by the corresponding anomalous  magnetic
moments: $\kappa^u=1.35$ and $\kappa^d=-1.80$ with
$\kappa^u/\kappa^d=0.75$. The experimental data 
for the proton and neutron anomalous magnetic moments, the $SU(2)$
approximation being considered, gives phenomenological values as
$\kappa^u=1.67$ and $\kappa^d=-2.03$ with $\kappa^u/\kappa^d=0.82$.  
The anomalous magnetic moment for the proton is obtained as
$\kappa_p=1.49$, which is about $17\,\%$ underestimated compared to
the experimental data $\kappa_p^{\mathrm{Exp}}=1.79$. The Pauli
form factor for the up quark falls off faster than that
for the down quark. Figure~\ref{fig:3} shows the proton
anomalous tensor magnetic form factors for the up and down quarks,
respectively. The up form factor also decreases faster than the down
form factor as $Q^2$ increases. We will see later that this difference
will clearly appear in the transverse quark spin densities. 
In general, the $\chi$QSM shows the tendency that the
slopes of the up quark form factors decrease faster than those of the
down quark form factors. This implies that the up quark radii are
smaller than those of the down quark so that the up quark seems to be
located more to the center of the proton than the down quark. 

\begin{table}[h]
\caption{Parameters of the proton up and down form factors fitted by
  the form $F(Q^{2})=F_{0}/(1+Q^{2}/m_p^{2})^{p}$. The charges of the
  quarks are excluded and the number of valence quarks are included for
  $F_{1}$ and $F_{2}$.}  
\label{tab:1}

\begin{tabular}{c|cccccc}
\hline \hline
 & $F_{1}^{u}$ & $F_{2}^{u}$ & $\kappa_{T}^{u}$ & $F_{1}^{d}$ &
 $F_{2}^{d}$ & $\kappa_{T}^{d}$\tabularnewline \hline 
$p$ & $2.72$ & $2.65$ & $2.43$ & $5.62$ & $3.02$ & $5.70$\tabularnewline
$Q^{2}=0$ & $2$ & $1.35$ & $3.56$ & $1$ & $-1.80$ & $1.83$\tabularnewline
$m_p$[GeV] & $1.07$ & $0.96$ & $0.97$ & $1.65$ & $1.12$ & $2.03$\tabularnewline
\hline\hline
\end{tabular}
\end{table}

In the lattice calculation~\cite{Gockeler:2006zu}, a simple $p$-pole
parametrization for the tensor form factors 
\begin{equation}
F(Q^2) = \frac{F_0}{(1+Q^2/m_p^2)^p}
\label{eq:p-pol}
\end{equation}
was employed with the three parameters $F_0=F(t=0)$, $m_p$, and $p$
for each form factor, where $F_0$ denotes the corresponding value of a
form factor at $Q^2=0$ and $m_p$ represents the $p$-pole mass in the
unit of GeV. Similarly, we also utilize the parametrization of
Eq.(\ref{eq:p-pol}) for the vector and tensor form factors.  
In Table~\ref{tab:1}, we list the numerical results of
the parameters for the parametrization for each form factor. 

Figure~\ref{fig:3} draws the proton Dirac and Pauli form factor
together with the anomalous tensor magnetic form factors of the
strange quark. The strange quark Dirac form factor is naturally equal
to zero at $Q^2=0$. The value of the strange quark Pauli form factor
is the same as the strange magnetic moment of the nucleon: 
$\mu_s=0.12\,\mu_N$~\cite{Silva:2001st}. The strange quark ATMM is
almost compatible with zero.  

Both the strange quark Dirac and anomalous tensor magnetic form
factors show $Q^2$ dependencies very similar to the electric form
factor of the neutron.  That is, they start to increase and then fall
off slowly. On the contrary, the strange Pauli form factor decreases
rather strongly till around $Q^2\approx 0.25\,\mathrm{GeV}^2$ and as
$Q^2$ increases, $F_2^s$ gets larger. Moreover, $F_2^s$ becomes
negative from around $Q^2\approx0.2\,\mathrm{GeV}^2$. However, the
general shape, multiplying the Pauli form factor by $-1$ and shifting
it by a constant, is the same for all three strange quark form
factors.  

Because of this $Q^2$ behavior, it is not as simple as the case of the
up and down form factors to parametrize the strange form
factors. Thus, we introduce the following parametrization for the
Fourier transform 
\begin{equation}
 F(Q^{2})=\left(a+bQ^{2}\right)e^{-c (Q^2)d},
\end{equation}
where the four parameters $a$, $b$, $c$, and $d$ are fitted to the form
factors obtained from the $\chi$QSM. The corresponding values for each
form factor are listed in Table~\ref{tab:2}.
\begin{table}[h]
\caption{Parameters of the proton strange quark form factors fitted in
the form of $F(Q^{2})=\left(a+bQ^{2}\right)e^{-c Q^{2d}}$. Note that
the charge of the strange quark is excluded in the case of $F_{1}$ and
$F_{2}$.}  
\label{tab:2}
\begin{tabular}{c|cccc}
\hline \hline
 & $a$ & $b$ & $c$ & $d$\tabularnewline
\hline 
$F_{1}^{s}$ & $0$ & $1.02$ & $2.95$ & $0.72$\tabularnewline
$F_{2}^{s}$ & $0.12$ & $-0.63$ & $2.59$ & $0.81$\tabularnewline
$\kappa_{T}^{s}$ & $0.01$ & $2.85$ & $2.87$ & $0.62$\tabularnewline
\hline\hline
\end{tabular}
\end{table}

\section{Results and discussion}
We now proceed to compute the transverse quark spin
densities in a proton. We consider first unpolarized quarks in a
transversely polarized nucleon with nucleon polarization $\bm S = (1,0)$ and
quark polarization $\bm s =(0,0)$. We express these polarizations in the
notation of $[\bm S,\bm s]= [(1,0),(0,0)]$. Afterwards we consider the
unpolarized nucleon and transveresly polarized quarks with $[\bm S,\,\bm
  s]=[(0,0),\,(1,0)]$. For these cases, the second line of Eq.(\ref{eq:rho}) 
does not contribute to the spin densities. 

\begin{figure}[ht]
\begin{center}
\includegraphics[scale=0.43]{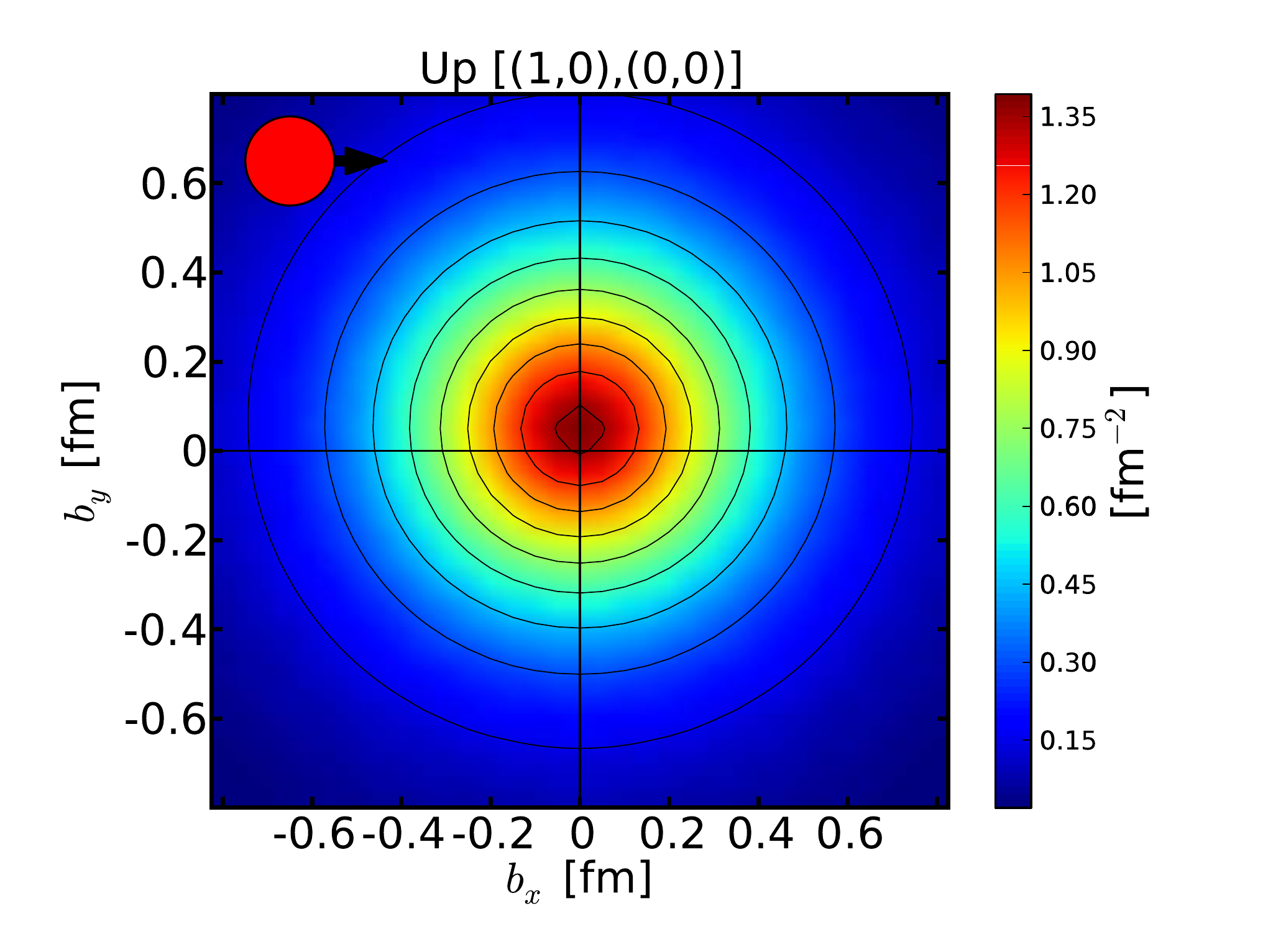}\includegraphics[scale=0.43]{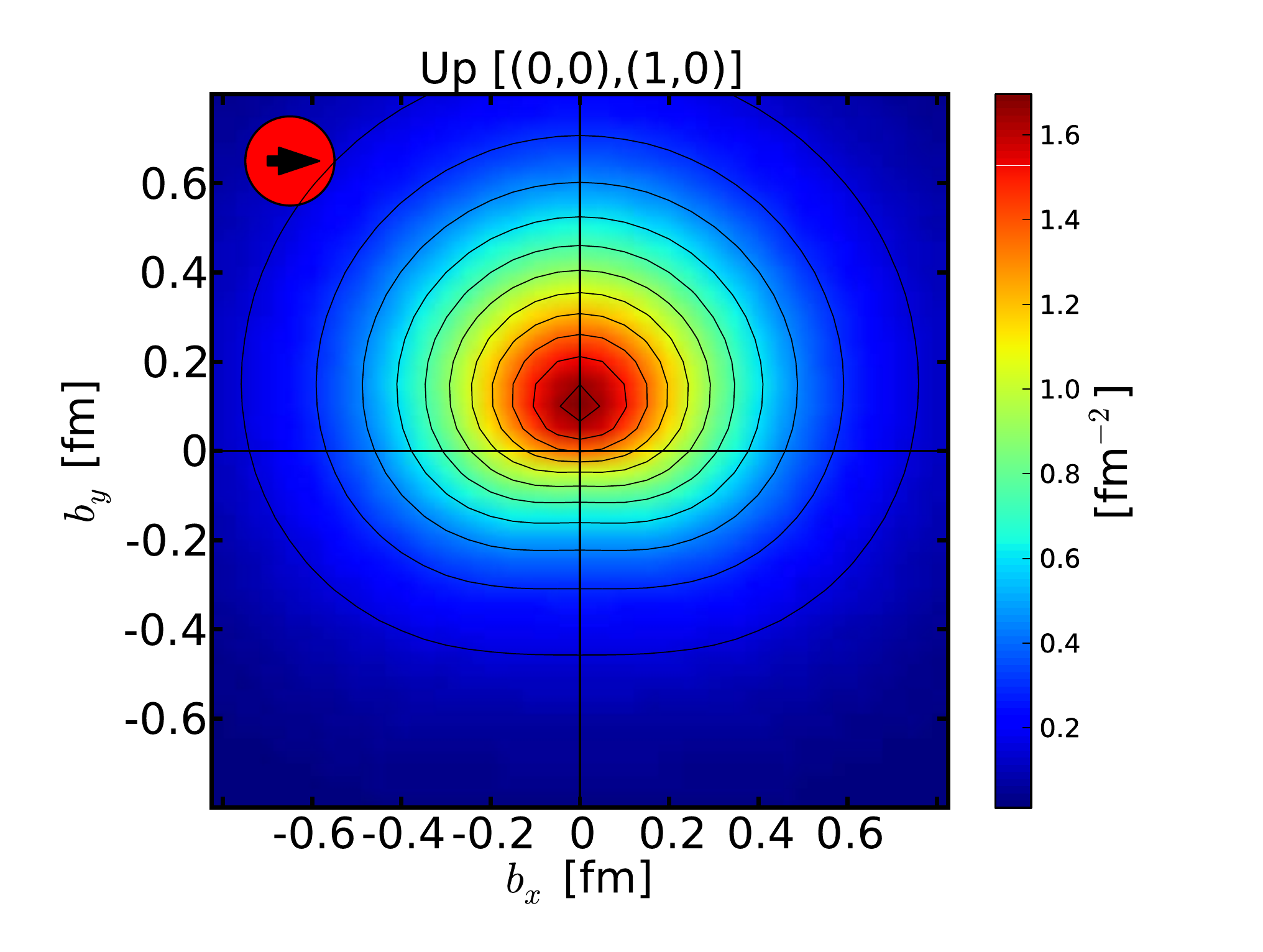}\\
\includegraphics[scale=0.43]{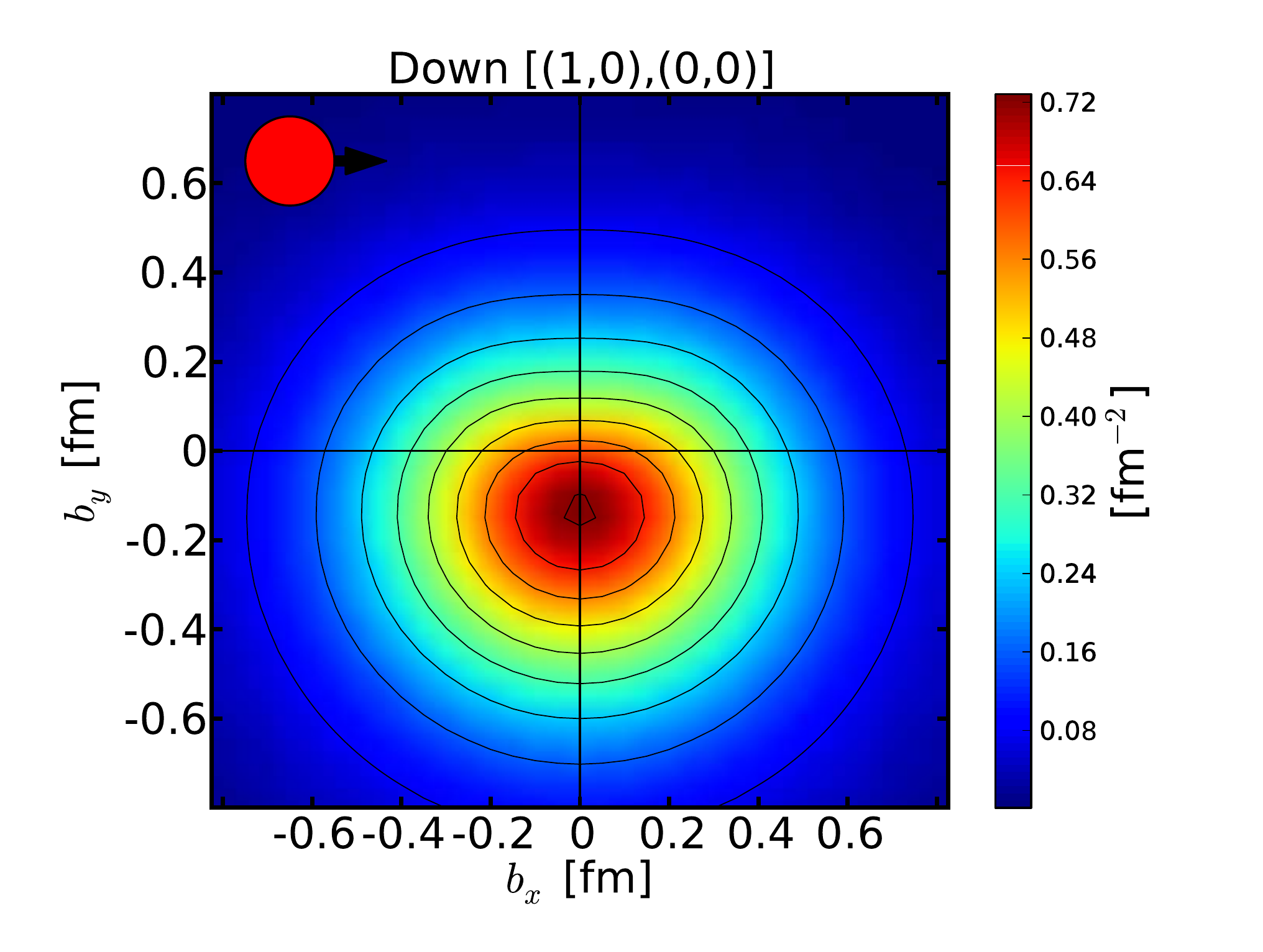}\includegraphics[scale=0.43]{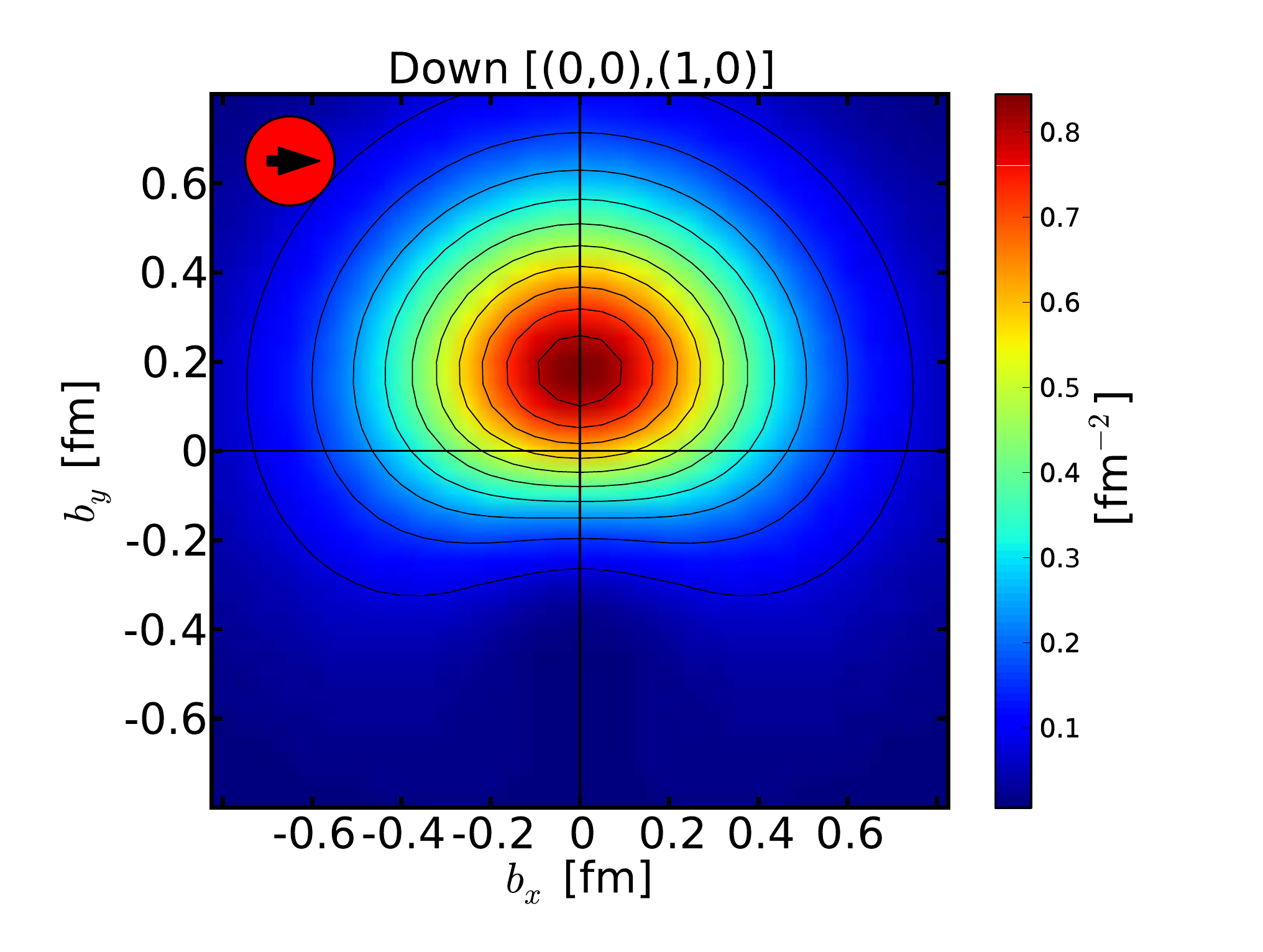}
\end{center}
\caption{Transverse up and down quark spin densities with the lowest
  moment of the nucleon from the $\chi$QSM.  In the upper left panel,
  the density of unpolarized quarks in a transversely polarized
  nucleon ($[\bm S,\,\bm s]=[(1,0),\,(0,0)]$) is drawn and in the
  upper right panel, that of transversely polarized quarks in a
  unpolarized nucleon ($[\bm S,\,\bm s]=[(0,0),\,(1,0)]$). In the
  lower panel, we plot the down quark densities.}
\label{fig:4}
\end{figure}
In Fig.~\ref{fig:4}, we illustrate the transverse up and down quark
spin densities of the nucleon. In the left panel we show the density 
of unpolarized up and down quarks in a transversely polarized nucleon
for $[\bm S,\,\bm s]=[(1,0),\,(0,0)]$. As can be seen in
Eq.(\ref{eq:rho}), the deformation of these densities are
governed by the Pauli form factors. We see that the down
quark density is more distorted in the negative direction of
$b_y$ than the up quark density in the positive direction. The reason
can be found in the fact that firstly the down Pauli form factor falls
off more slowly than that of the up quark and secondly, the down
anomalous magnetic moment ($\kappa^d=-1.80$) is negative. The sign of
the form factors at intermediate $Q^2$ determines the direction of the
shift. In the case of choosing the polarizations as  $[\bm S,\,\bm
s]=[(0,0),\,(1,0)]$, only the anomalous tensor magnetic
form factors contribute, as shown in the right panel of
Fig.~\ref{fig:4}. Because both $\kappa_T^u$ and $\kappa_T^d$ are
positive, both transverse spin densities are deformed in the direction
of positive $b_y$ and again, because the form factor $\kappa_T^d$ falls
off more slowly, the density for the down quark is more strongly
deformed. Furthermore, we want to mention that our results for the
transverse up and down quark spin densities of the nucleon with the
lowest moment are very similar to those from the lattice 
calculation~\cite{Gockeler:2006zu}. 

\begin{figure}[ht]
\begin{center}
\includegraphics[scale=0.43]{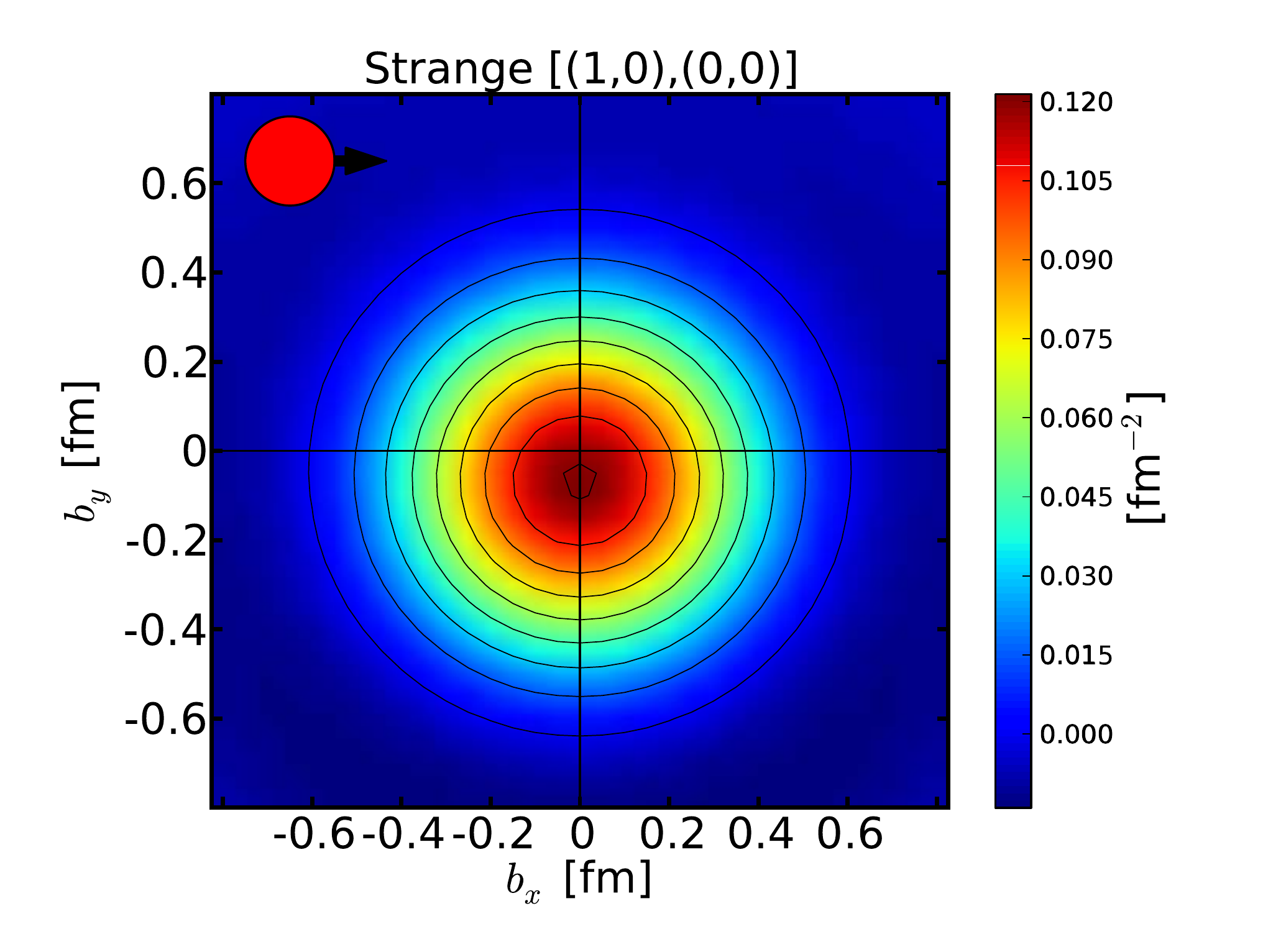}
\includegraphics[scale=0.43]{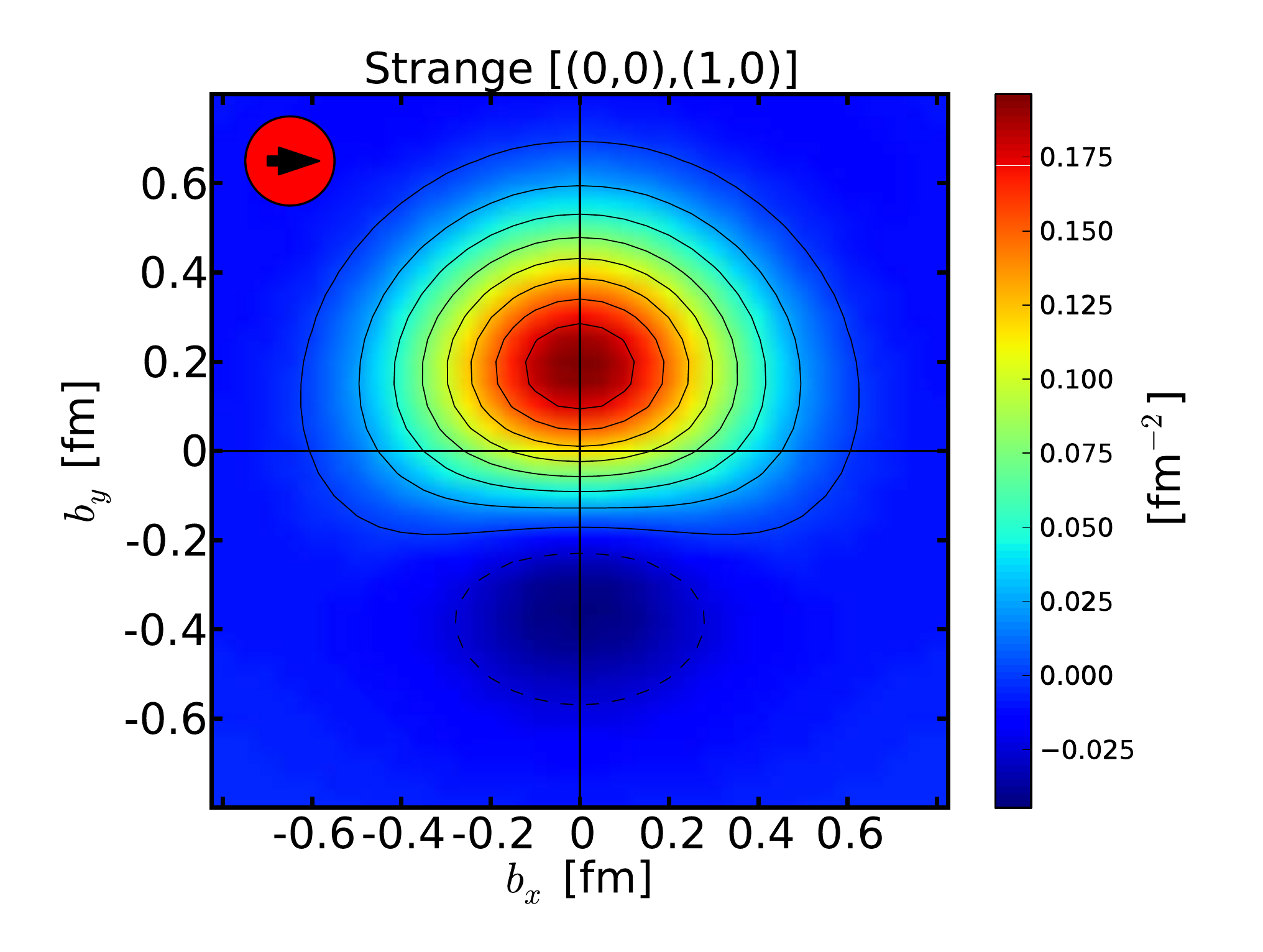}
\end{center}
\caption{Transverse strange quark spin densities of the nucleon
  from the $\chi$QSM with the lowest moment.  In the left panel,
  the density of unpolarized strange quarks in a transversely polarized
  nucleon ($[\bm S,\,\bm s]=[(1,0),\,(0,0)]$) is drawn and in the right panel,
  that of transversely polarized strange quarks in an unpolarized
  nucleon ($[\bm S,\,\bm s]=[(0,0),\,(1,0)]$).} 
\label{fig:5}
\end{figure}
Since both the strange Pauli form factor and anomalous tensor magnetic
form factor turn out to be rather small, we can expect that the
strength of the strange densities will be also quite small.
Nevertheless, it is of great interest to see how much the transverse
strange densities are distorted. Figure~\ref{fig:5} plots the
transverse strange quark spin densities with the lowest moment.  Note
that the magnitudes of the densities are smaller than those of the up
and down quarks by an order of magnitude.  

\begin{figure}[ht]
\begin{center}
\includegraphics[scale=0.4]{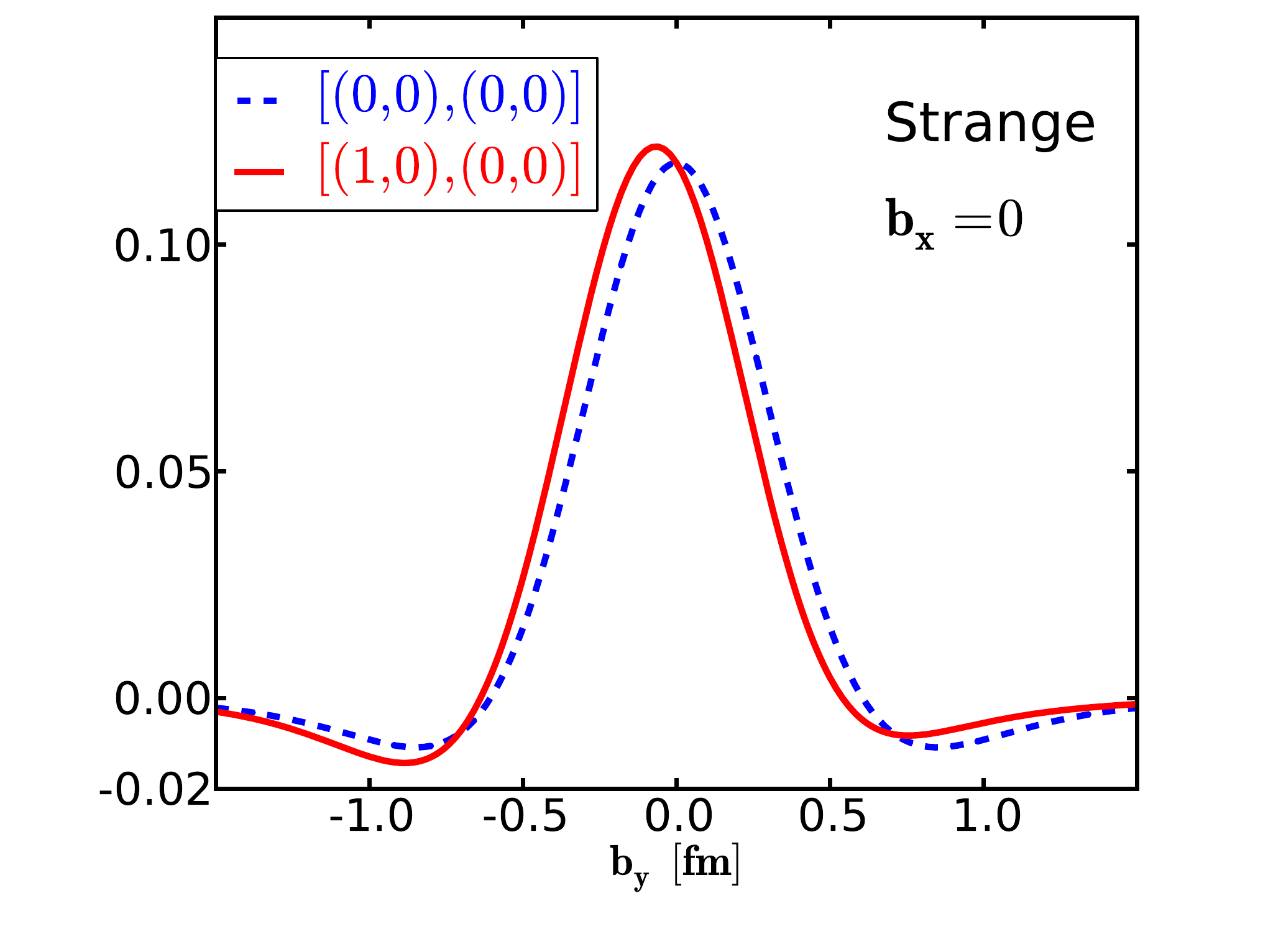}
\includegraphics[scale=0.4]{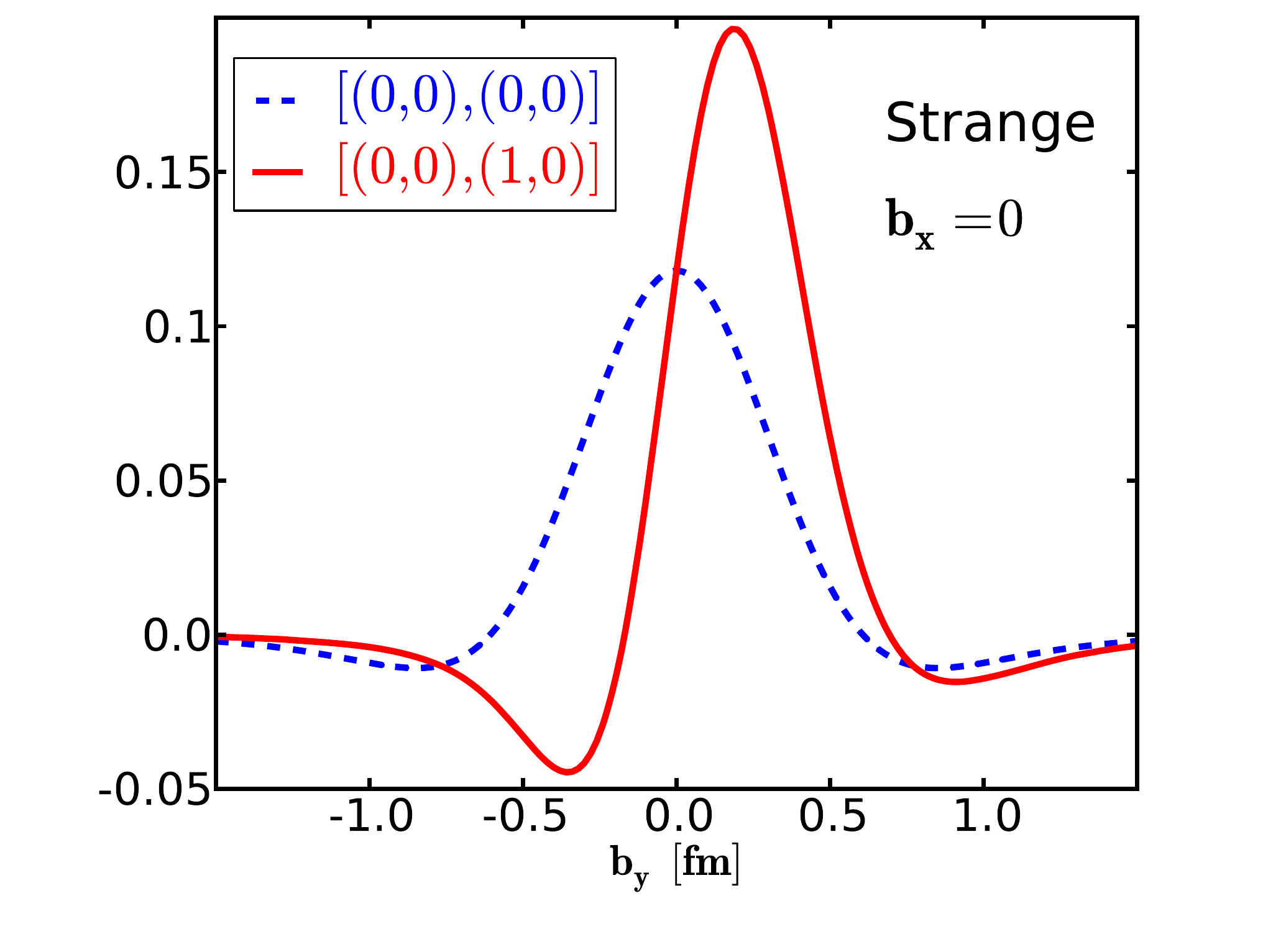}
\end{center}
\caption{Transverse strange quark spin densities of the nucleon
  from the $\chi$QSM with the lowest moment with $b_x=0$. In the left
  panel, the density of unpolarized strange quarks in a transversely
  polarized nucleon ($[\bm S,\,\bm s]=[(1,0),\,(0,0)]$) is drawn and
  in the right panel, that of transversely polarized strange quarks in
  an unpolarized nucleon ($[\bm S,\,\bm s]=[(0,0),\,(1,0)]$).}   
\label{fig:7}
\end{figure}

It is interesting to see that the density of unpolarized strange
quarks in a polarized nucleon is negatively shifted. It is due to the
fact that the strange Pauli form factor $F_2^s$ turns negative from
$Q^2\approx 0.2\mathrm{GeV}^2$ and the lower $Q^2$ values are suppressed in
the Fourier transform. Therefore, the negative values of the form
factor at intermediate $Q^2$ shift the density to the negative $b_y$
direction, despite the positive strange anomalous magnetic moment. On
the contrary, the density of polarized strange quarks in an unpolarized
nucleon is shifted and remarkably distorted in the direction of the
positive $b_y$.  This can be understood from the $Q^2$ dependence of 
$\kappa_T^s(Q^2)$ as shown in Fig.~\ref{fig:3}. Moreover, the density
becomes negative for the negative values of $b_y$. One can see later
this more clearly (see Fig.~\ref{fig:7}).

\begin{figure}[ht]
\begin{center}
\includegraphics[scale=0.4]{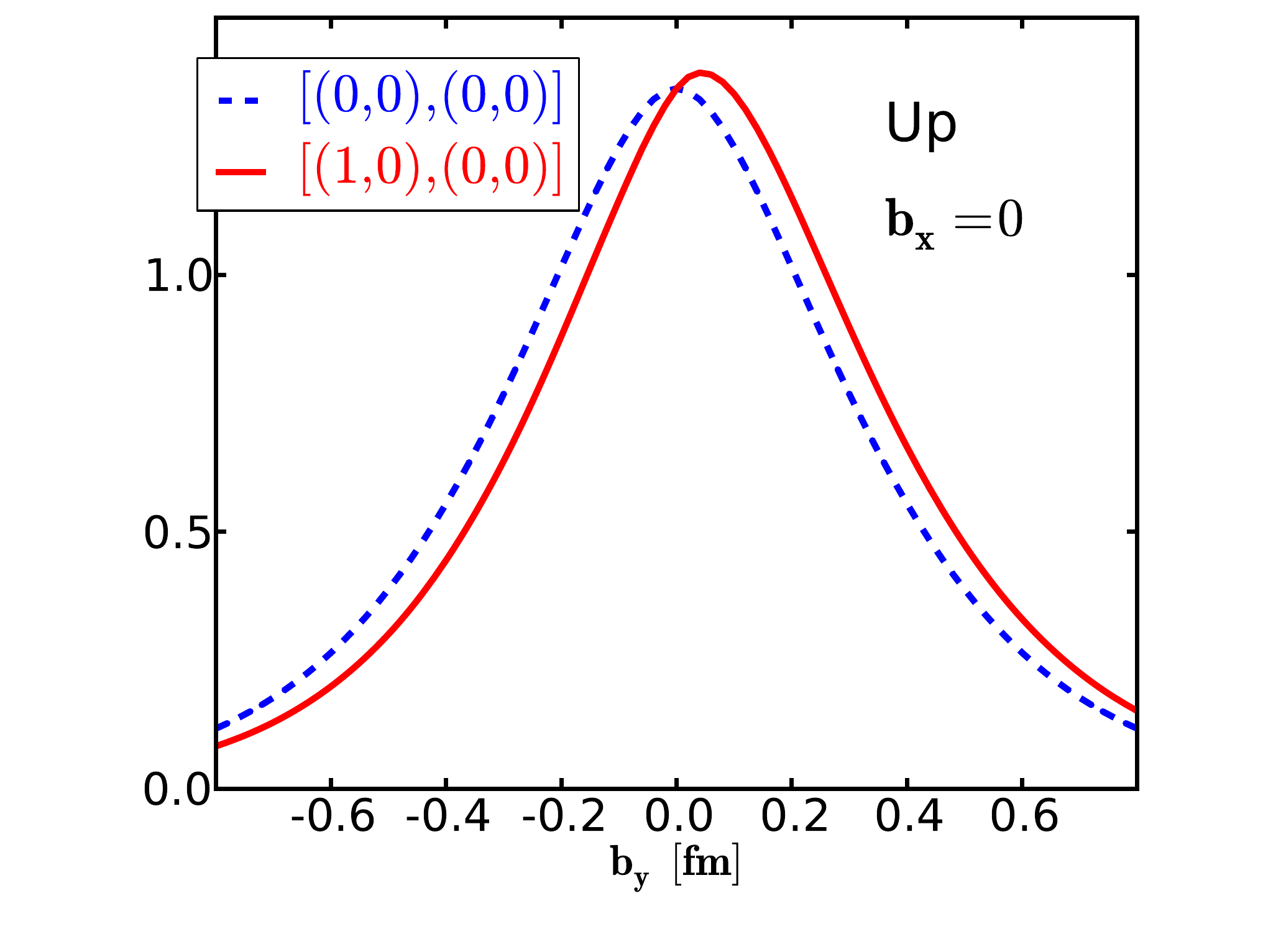}\includegraphics[scale=0.4]{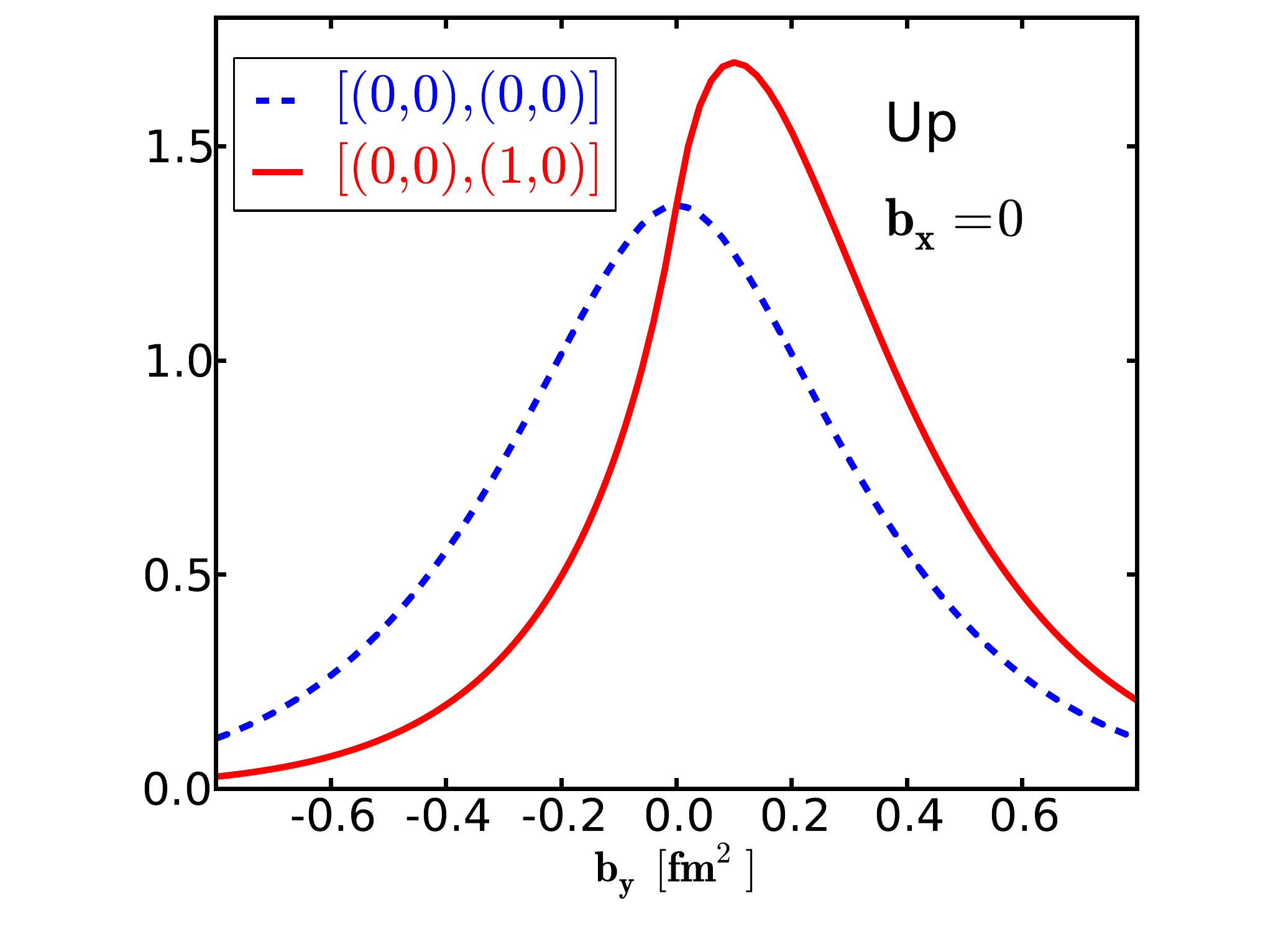}\\
\includegraphics[scale=0.4]{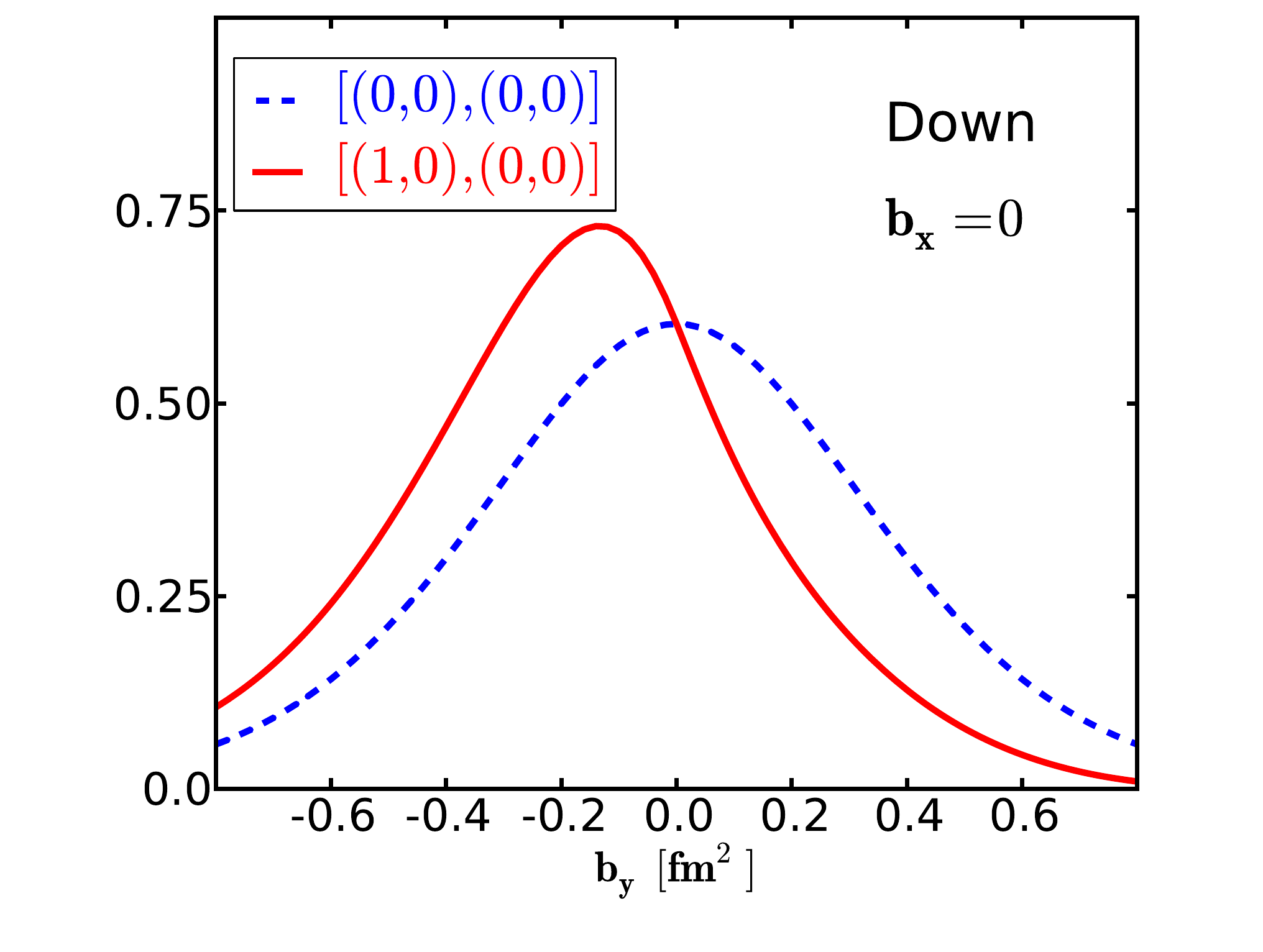}\includegraphics[scale=0.4]{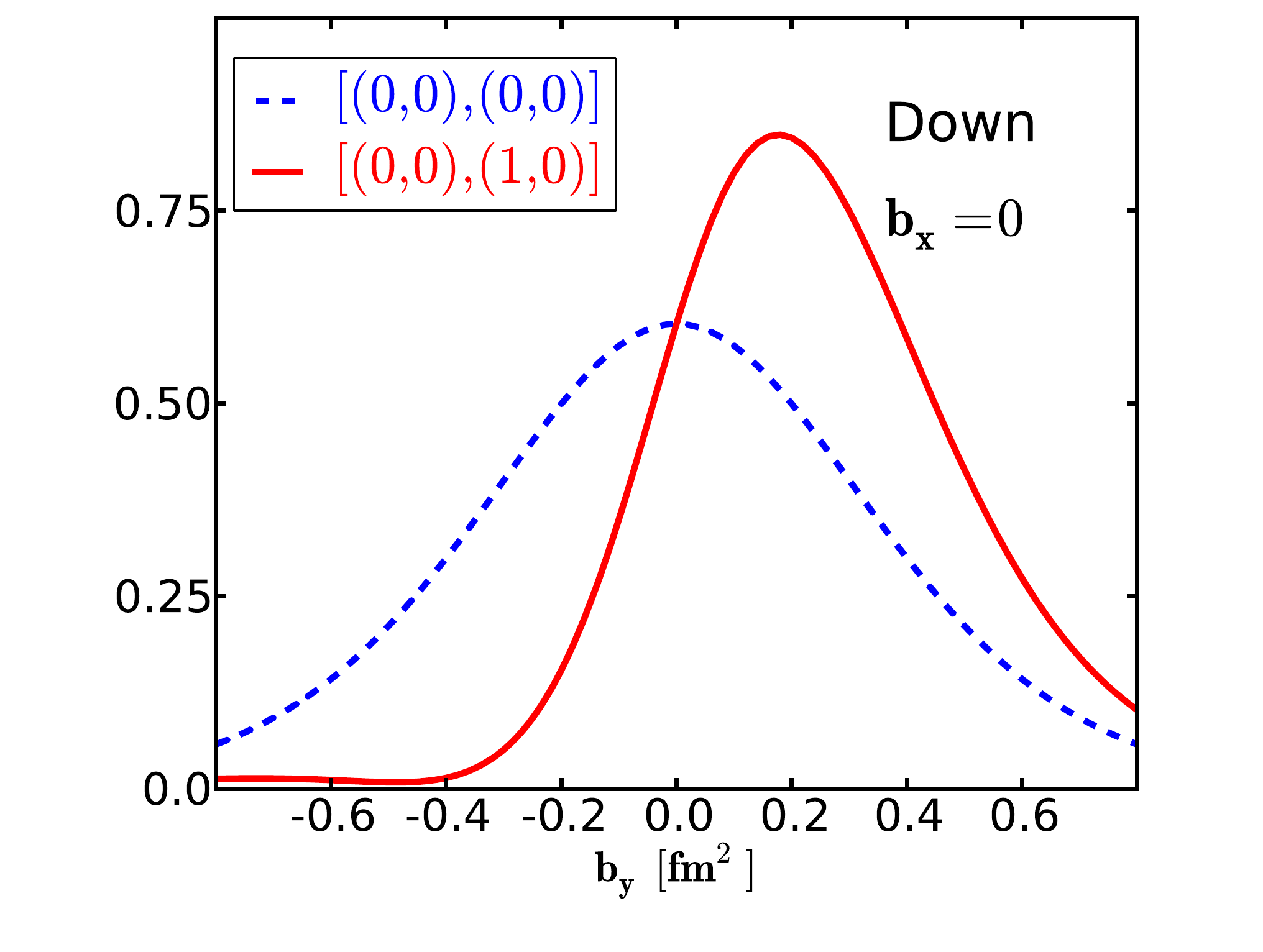}
\end{center}
\caption{Transverse up and down quark spin densities with the lowest moment 
of the nucleon from the $\chi$QSM with $b_x=0$.  In the upper left panel, 
  the density of unpolarized quarks in a transversely polarized
  nucleon ($[\bm S,\,\bm s]=[(1,0),\,(0,0)]$) is drawn and in the
  upper right panel, that of transversely polarized quarks in a
  unpolarized nucleon ($[\bm S,\,\bm s]=[(0,0),\,(1,0)]$). In the
  lower panel, we plot the down quark densities.}
\label{fig:6}
\end{figure}

In Fig.~\ref{fig:6} we draw the transverse up and down quark spin
densities with $b_x$ fixed to be zero. In the case of the unpolarized
up quarks in a polarized nucleon, the density is slightly shifted to
the positive direction of $b_y$, whereas that of the polarized up
quarks in an unpolarized nucleon is quite much shifted to the positive
$b_y$ direction. Moreover, the width of the profile gets narrower and
more peaked. On the contrary, the transverse down quark spin density 
for the unpolarized quarks in a polarized nucleon is changed to the
negative $b_y$ direction and shows an obvious distortion. That for the
polarized quarks in an unpolarized nucleon is shifted to the positive
$b_y$ direction and distorted again. 

Figure~\ref{fig:7} shows the profiles of the transverse strange quark
spin densities with $b_x=0$. Interestingly, the density for the
unpolarized strange quarks in a polarized nucleon is only slightly
modified. In contrast, the polarized strange quarks are strongly
redistributed in an unpolarized nucleon. As a result, the peak
position is shifted to the positive $b_y$ direction and becomes
sharper. Moreover, the density becomes even more negative for the
negative value of $b_y$ than for the positive $b_y$. 

\section{Summary and conclusion}
We investigated the transverse quark spin densities of the
nucleon with the lowest moment, using the results of the vector and
tensor form factors derived from the SU(3) chiral quark-soliton
model. we first recapitulated the flavor-decomposed anomalous vector
and tensor magnetic form factors as functions of the momentum
transfer. For numerical convenience, we made parametrizations of the
form factors. Having combined these form factors and performed the
Fourier transformations, we evaluated the transverse quark spin
densities of the nucleon with the lowest moment. We considered two
different cases, the density of unpolarized quarks in a polarized
nucleon and that of polarized quarks in an unpolarized nucleon. The
results turned out to be rather similar to the lattice QCD ones.
Transversly polarized quarks in an unpolarized proton are both shifted
to the positive direction of $b_x$. The shift is more prominent than
the one occuring for unpolarized quarks in a polarized proton, where
the density for the u-quark is shifted to positive and that of the
d-quark to the negative $b_y$ direction. 

We presented in this work the first result of the transverse strange
quark spin densities. Since the magnitudes of strange Pauli and
anomalous tensor magnetic form factors are rather small, the strange
densities turn out to be much smaller than those for the up and down
quarks. However, the density for polarized strange quarks in an
unpolarized nucleon is noticeably distorted in the direction of the
positive $b_y$ and becomes more negative for the negative values of
$b_y$.  

In the present work, we considered only the transverse spin
densities with the lowest moment. The generalized form factors with
higher moments can be calculated in principle within the framework of
the chiral quark-soliton model. However, we need to consider them
carefully because of the presence of the derivative operators. The
corresponding work is under progress. So far, we did not consider
other cases of the transverse polarizations such as
$[(1,0),\,(1,0)]$ for which one requires information on the form
factor $\tilde{H}_T$ and its derivative. However, we found that
this form factor showed a numerically sensitivity. A related work
addressing that is under way.

\begin{acknowledgments}
H.Ch.K is grateful to E. Epelbaum and M.V. Polyakov for hospitalities
during his visit to Theoretische Physik Institute II in
Ruhr-Universit\"at Bochum, where part of the work was carried out. 
The authors are grateful to A. Metz for valuable comments and
suggestions. The present work was supported by Basic Science Research
Program through the National Research Foundation of Korea (NRF) funded
by the Ministry of Education, Science and Technology (grant number: 
2010-0016265).    
\end{acknowledgments}

\end{document}